# Diversity-Multiplexing Tradeoff in the Low-SNR Regime

Sergey Loyka and Georgy Levin

*Abstract*—An extension of the popular diversity-multiplexing tradeoff framework to the low-SNR (or wideband) regime is proposed. The concept of diversity gain is shown to be redundant in this regime since the outage probability is SNR-independent and depends on the multiplexing gain and the channel power gain statistics only. The outage probability under the DMT framework is obtained in an explicit, closed form for a broad class of channels. The low and high-SNR regime boundaries are explicitly determined for the scalar Rayleigh-fading channel, indicating a significant limitation of the SNR-asymptotic DMT when the multiplexing gain is small.

## I. INTRODUCTION

SINCE error rate or outage probability analysis of MIMO channels/systems is a challenging task, an elegant framework termed "diversity-multiplexing tradeoff" has been proposed by Zheng and Tse [1], which quantifies the error rate or outage performance at the high SNR regime via the two principal gains offered by a MIMO channel: diversity and multiplexing gains. Because this SNR-asymptotic analysis is feasible for many systems/channels, which resisted successfully other lines of attacks otherwise, the DMT framework became very popular and have been successfully applied to many systems/channels/space-time code designs [2][3], despite some difficulties due to the asymptotic nature of the analysis [4]. The SNR-asymptotic DMT of [1] has been extended to finite SNR in [4].

It is a common belief that the DMT framework applies only in the high-SNR regime. In this Letter, we demonstrate that this is not so: the DMT framework is also applicable in the low-SNR regime (also known as the wideband regime [7]), at which many practical systems (e.g. CDMA) operate [1] [2][7]. Only a minor modification to the original framework is required to accomplish this. Furthermore, unlike the high-SNR regime, the channel outage probability can be found in an explicit closed-form (via the channel power gain), which is independent of the SNR so that the concept of diversity gain becomes redundant at low SNR (but the concept of multiplexing gain is very much relevant, albeit in a slightly modified form).

## II. SYSTEM MODEL AND HIGH-SNR DMT

Let us consider a frequency-flat, quasi-static (block-fading) MIMO channel,

$$\mathbf{r} = \mathbf{H}\mathbf{s} + \xi \quad (1)$$

[1] while low SNR implies high error rate in uncoded systems, it is not so in coded ones e.g. many CDMA systems operate in low-SNR, low error rate regime [2].

where $\mathbf{s}$ and $\mathbf{r}$ are the Tx and Rx vectors correspondingly, $\mathbf{H}$ is the $n \times m$ channel matrix, i.e. the matrix of the complex channel gains between each Tx and each Rx antenna, $m$ and $n$ are the numbers of Tx and Rx antennas, and $\xi$ is the additive white Gaussian noise (AWGN), which is assumed to be $\mathcal{CN}(\mathbf{0}, \sigma_0^2 \mathbf{I})$, i.e. independent and identically distributed (i.i.d.) in each branch. When full channel state information (CSI) is available at the Rx end but no CSI or its distribution at the Tx end, isotropic signalling is optimal [5] and the instantaneous channel capacity (i.e. the capacity of a given channel realization $\mathbf{H}$) in nats/s/Hz is given by the celebrated log-det formula [9][10],

$$C = \ln \det \left( \mathbf{I} + \frac{\gamma}{m} \mathbf{H}\mathbf{H}^+ \right) \quad (2)$$

where $\gamma$ is the average SNR per Rx antenna (contributed by all Tx antennas), "+" denotes conjugate transpose. Without loss of generality, we assume that the channel matrix is normalized, $E \|\mathbf{H}\|_F^2 = \sum_{i,j} E |h_{ij}|^2 = mn$, where $E$ denotes expectation (over fading) and $h_{ij}$ are the entries of the channel matrix $\mathbf{H}$.

The channel outage probability $P_{out}$ is the probability that the fading channel is not able to support the target transmission rate $R$,

$$P_{out} = \Pr\{C < R\} \quad (3)$$

Defining the multiplexing gain $r$ as

$$r = \lim_{\gamma \to \infty} R/\ln \gamma \quad (4)$$

and the diversity gain as[2]

$$d = -\lim_{\gamma \to \infty} \ln P_{out} / \ln \gamma \quad (5)$$

the SNR-asymptotic ($\gamma \to \infty$) DMT for the independent identically distributed (i.i.d.) Rayleigh fading channel with the coherence time in symbols $T \geq m$ can be compactly expressed as [1][3],

$$d(r) = (n-r)(m-r), \ r = 0, 1, \ldots \min(m, n) \quad (6)$$

where $m, n$ are the number of transmit (Tx), receive (Rx) antennas, for integer values of $r$, and using the linear interpolation in-between. This result has been also extended to many other scenarios (see e.g. the references in [4]). We note that the motivation for the definition of $r$ in (4) is that the mean (ergodic) capacity $\overline{C}$ scales as $\min(m,n) \ln \gamma$ at high SNR, $\overline{C} \approx \min(m,n) \ln \gamma$ as $\gamma \to \infty$, and the motivation for the definition of $d$ in (5) is that $P_{out}$ scales as $\gamma^{-d}$ at high SNR,

$$P_{out} \approx c/\gamma^d, \text{ as } \gamma \to \infty \quad (7)$$

[2] while the original definition in [1] employed the average error rate, we use the channel outage probability instead since it is the best achievable average error rate [6], and average error rate in general is dominated by the outage probability in the low outage regime [2].

2where $c$ is a constant independent of the SNR. While this constant is not a part of the DMT framework, it affects significantly the outage probability and its value can be found from the size-asymptotic theory [4].

## III. DIVERSITY-MULTIPLEXING TRADEOFF IN THE LOW-SNR REGIME

At finite SNR, including the low-SNR regime, the multiplexing gain definition in (4) has to be properly modified [4],

$$R = r \ln(1+\gamma) \stackrel{(a)}{\approx} r\gamma \qquad (8)$$

i.e. it defines the target rate $R$ as a fraction of the non-fading AWGN channel capacity $\ln(1+\gamma)$ and (a) holds in the low-SNR regime $\gamma \ll 1$ (from (8), it is also the wideband regime $R \ll 1$ [7] when $r$ is not too large). Note that this definition is suitable for rate-adaptive systems, where the rate scales as a function of the average SNR. We are now in a position to characterize the channel outage probability in the low-SNR or wideband regime under the DMT framework.

**Proposition 1.** *At the low-SNR regime $\gamma \ll 1$, the channel outage probability for the target rate $R$ in (8) can be approximated as*

$$P_{out}(r) \approx F_H(mr) \qquad (9)$$

*where $F_H(x) = \Pr[\|\mathbf{H}\|_F^2 < x]$ is the cumulative distribution function (CDF) of the channel power gain $\|\mathbf{H}\|_F^2$.*

*Proof:* From the definition of the outage probability in (3) and (2), one obtains

$$\begin{aligned} P_{out} &= \Pr\left[\ln\det\left(\mathbf{I} + \frac{\gamma}{m}\mathbf{H}\mathbf{H}^+\right) < R\right] \\ &= \Pr\left[\sum_{i=1}^{\min(m,n)} \ln\left(1 + \frac{\gamma}{m}\lambda_i\right) < R\right] \\ &\stackrel{(a)}{\approx} \Pr\left[\frac{\gamma}{m}\sum_{i=1}^{\min(m,n)} \lambda_i < R\right] \\ &\stackrel{(b)}{=} \Pr\left[\|\mathbf{H}\|_F^2 < \frac{mR}{\gamma}\right] \\ &\stackrel{(c)}{\approx} F_H(mr) \end{aligned} \qquad (10)$$

where $\lambda_i$ are the eigenvalues of $\mathbf{HH}^+$; (a) follows from the fact that $\ln(1+x) \approx x$ for $x \ll 1$ (which applies under the low-SNR condition $R \approx r\gamma \ll 1$), (b) follows from the fact that $\sum_i \lambda_i = tr(\mathbf{HH}^+) = \|\mathbf{H}\|_F^2$, and (c) follows from (8). ∎

Note that the outage probability is not a function of the SNR at the low-SNR regime but depends only on the multiplexing gain and the statistics of the channel power gain (for any fading distribution). Therefore, the concept of diversity gain becomes redundant at this regime and thus the DMT-based design of space-time codes (see e.g. [3]) is not necessarily optimal anymore. In fact, since the outage probability is known in closed form (see e.g. Corollaries 1-3 below), we have arrived to the "outage probability-multiplexing gain tradeoff". This tradeoff inherits all the properties of $F_H(\cdot)$ and should be used as a basis for code design of rate-adaptive systems in the low SNR/wideband regime.

Since $E\|\mathbf{H}\|_F^2 = mn$, it follows from (10) that the outage probability is low ($P_{out} < 1/2$) if $r < n$, i.e. the multiplexing gain should not exceed the number of receive antennas.

We remark that while the outage probability analysis of a MIMO channel at arbitrary or even high SNR is a formidable analytical task (compare (10) to the analysis in [1]), the low-SNR case is much easier, since no joint eigenvalue density is required but rather the distribution of the (scalar) channel power gain is sufficient, which is known in a closed form for many cases. We consider those below.

**Corollary 1.** *In the i.i.d Rayleigh-fading channel, the outage probability under the DMT framework at the low-SNR regime in (9) becomes*

$$P_{out}(r) \approx F_{mn}(mr) \approx \frac{(mr)^{mn}}{(mn)!} \qquad (11)$$

*where $F_k(x) = 1 - e^{-x}\sum_{i=0}^{k-1} x^i/i!$ is the outage probability of $k$-th order maximum ratio combiner over the i.i.d. Rayleigh-fading channel, and $2^{nd}$ equality holds in the low outage regime, $mr \ll 1$.*

*Proof:* By observing that the CDF of $\|\mathbf{H}\|_F^2$ is $F_{mn}(x)$ in this case. ∎

We note that the outage probability in (11) is a monotonically-decreasing function of $m, n$ and a monotonically-increasing function of $r$.

**Corollary 2.** *In the arbitrary-correlated Rayleigh-fading channel, the outage probability under the DMT-framework in the low-SNR regime is*

$$P_{out}(r) \stackrel{(a)}{\approx} 1 - \sum_{i=1}^{mn} A_i \exp\left(-\frac{mr}{\lambda_i}\right) \stackrel{(b)}{\approx} \frac{1}{\det \mathbf{R}} \frac{(mr)^{mn}}{(mn)!} \qquad (12)$$

*where $\lambda_i$ are the (distinct) eigenvalues of the channel correlation matrix $\mathbf{R} = E[vec(\mathbf{H})vec(\mathbf{H})^+]$, $vec(\mathbf{H})$ denotes column-wise vectorization, $E$ denotes expectation over the channel statistics, $A_k = \prod_{i \neq k} \lambda_k/(\lambda_k - \lambda_i)$ are the partial fraction decomposition coefficients; (b) holds in the low-outage regime $mr \ll 1$ provided that $\mathbf{R}$ is non-singular (if it is, only non-zero eigenvalues should be retained in (12)).*

*Proof:* (a) follows from the fact that the CDF of $\|\mathbf{H}\|_F^2 = |vec(\mathbf{H})|^2$ is $1 - \sum_{i=1}^{mn} A_i \exp(-x/\lambda_i)$ (see e.g. Appendix A in [8]). (b) follows from the low-outage approximation of this CDF. ∎

Note that Corollary 2 holds for arbitrary correlation structure of the channel, not only when it is Kronecker one. In the latter case, $\mathbf{R} = \mathbf{R}_t^T \otimes \mathbf{R}_r$ [8], where $\mathbf{R}_r = E[\mathbf{HH}^+]$, $\mathbf{R}_t = E[\mathbf{H}^+\mathbf{H}]$ are the Rx and Tx end correlation matrices, $\otimes$ denotes Kronecker product of two matrices, and (12) holds with the substitution $\lambda_i \rightarrow \lambda_{ij} = \eta_{ri}\eta_{tj}$, where $\eta_{ri}$, $\eta_{tj}$ are the eigenvalues of $\mathbf{R}_r$, $\mathbf{R}_t$ respectively.

**Corollary 3.** *Under the conditions of Corollary 2 and when the channel correlation has Kronecker structure,*

$$P_{out}(r) \approx \frac{1}{(\det \mathbf{R}_r)^m (\det \mathbf{R}_t)^n} \frac{(mr)^{mn}}{(mn)!} \qquad (13)$$

Note from (12), (13) that correlation has a clearly negative effect on the outage probability in the low-outage regime, which is measured by the determinant of the correlation matrices (since $\det \mathbf{R}$ is maximum when there is no correlation and is strictly smaller otherwise).

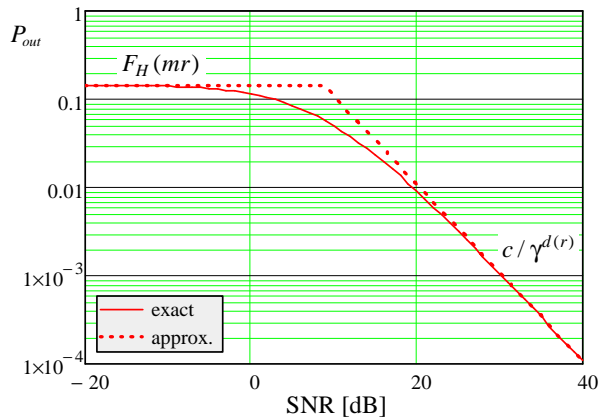

Fig. 1. Outage probability and its approximation for the $2 \times 2$ i.i.d. Rayleigh fading channel; $r = 1$, $d(r) = 1$. The approximate outage is from (14) and the exact one was obtained by integrating the Wishart eigenvalue density (see e.g. [9]). Note that the approximation in (14) is very accurate when either $\gamma \ll 1$ or $\gamma \gg 1$ and less accurate in the transition region.

Finally, based on the high and low-SNR approximations above, we propose a piece-wise linear (on log-log scale) approximation of the outage probability under the DMT framework over the whole SNR range,

$$P_{out}(r) \approx \min\left[F_H(mr),\ c/\gamma^{d(r)}\right] \quad (14)$$

which is sufficiently accurate when $\gamma \ll 1$ or $\gamma \gg 1$ and may be less accurate in the transition region. Fig.1 shows a typical example for a 2x2 i.i.d. Rayleigh-fading channel.

## IV. OUTAGE PROBABILITY AND DMT IN $1 \times 1$ RAYLEIGH-FADING CHANNEL

While the results above hold true for a broad class of fading channels, we consider here a textbook-type example of the scalar (1x1) Rayleigh fading channel to get some additional insight into the behavior of the outage probability over the whole SNR range under the DMT framework, not just in low or high-SNR regimes, and to establish their boundaries.

In this case, the instantaneous channel capacity in (2) becomes $C = \ln(1 + |h|^2 \gamma)$, where $h$ is a scalar channel gain, and the outage probability at arbitrary SNR is

$$P_{out}(r) = F_h\left(\frac{(1+\gamma)^r - 1}{\gamma}\right) \quad (15)$$

where $F_h(x) = 1 - e^{-x}$ is the CDF of the channel power gain $|h|^2$. The high-SNR regime corresponds to $\gamma^r \gg 1$, so that (15) simplifies to

$$P_{out}(r) \approx F_h(1/\gamma^{1-r}) \approx 1/\gamma^{1-r}, \quad 0 < r < 1, \quad (16)$$

and $d(r) = 1 - r$, as expected in this channel [2]. Note, however, that the high-SNR condition $\gamma^r \gg 1$ is, within 10% accuracy for the argument of $F_h$ in (16), $\gamma^r \geq 10$, so that

$$\gamma \geq 10^{1/r} \quad (17)$$

Clearly, the high-SNR boundary $10^{1/r}$ is practically-reasonable when $r$ is not too small, i.e. 20 dB for $r = 0.5$, but very quickly (exponentially) increases when $r$ approaches 0, i.e. 100 dB for $r = 0.1$. This demonstrates that the SNR-asymptotic DMT is practically-relevant when the multiplexing gain is not too small (say $r \geq 0.5$), but quickly becomes irrelevant when $r \to 0$. Similar conclusions have been obtained in a more general setting in [4]. This may have significant consequences for the SNR-asymptotic DMT-based design of space-time codes for small values of $r$ (e.g. as in [3]). On the other hand, the low-SNR regime corresponds to $\gamma \ll 1$, i.e. $\gamma \leq 0.1$ within 10% accuracy for any value of the multiplexing gain $r$, and (15) simplifies to

$$P_{out}(r) \approx F_h(r), \quad 0 \leq r \leq 1, \quad (18)$$

so that the low-SNR approximation is much more robust compared to the high-SNR one.

## V. CONCLUSION

The SNR-asymptotic diversity-multiplexing tradeoff has been extended to the low-SNR (or wideband) regime. In this regime, the concept of diversity gain becomes redundant as the outage probability does not depend on the SNR but is a function of the multiplexing gain and the channel power gain statistics only. In fact, the outage probability under the DMT framework is obtained in an explicit closed form in this regime and should be used as a design criterion for low-SNR (wideband) systems/codes instead of the SNR-asymptotic DMT. The low and high-SNR boundaries have been explicitly characterized for the $1 \times 1$ Rayleigh-fading channel, indicating a severe limitation of the SNR-asymptotic DMT when the multiplexing gain becomes small.

## REFERENCES

[1] L. Zheng, D. N. C. Tse, "Diversity and Multiplexing: A Fundamental Tradeoff in Multiple-Antenna Channels", *IEEE Trans. Inform. Theory*, vol. 49, no. 5, pp. 1073-1096, May 2003.
[2] D.N.C. Tse, P. Viswanath, *Fundamentals of Wireless Communications*, Cambridge University Press, 2005.
[3] P. Elia et al, Explicit space-time codes achieving the diversity-multiplexing gain tradeoff, IEEE Trans. Inf. Theory, vol. 52, pp. 3869–3884, Sep. 2006.
[4] S. Loyka, G. Levin, Finite-SNR Diversity-Multiplexing Tradeoff via Asymptotic Analysis of Large MIMO Systems, IEEE Transactions on Information Theory, v. 56, N. 10, pp. 4781-4792, Oct. 2010.
[5] D. Palomar, J.M. Cioffi, M.A. Lagunas, Uniform Power Allocation in MIMO Channels: A Game-Theoretic Approach, IEEE Trans. Information Theory, v. 49, N. 7, pp. 1707-1727, Jul. 2003.
[6] S. Verdu, T.S. Han, "A General Formula for Channel Capacity", *IEEE Transactions on Information Theory*, vol. 40, no. 4, pp. 1147-1157, July 1994.
[7] S. Verdu, "Recent Results on the Capacity of Wideband Channels in the Low-Power Regime," IEEE Wireless Communications, pp. 40–45, August 2002.
[8] G. Levin, S. Loyka, On the Outage Capacity Distribution of Correlated Keyhole MIMO Channels, IEEE Transactions on Information Theory, v. 54, N. 7, pp. 3232-3245, July 2008.
[9] I. E. Telatar, "Capacity of multi-antenna Gaussian channels," AT&T Bell Labs, Internal Tech. Memo, 1995.
[10] G. J. Foschini and M. J. Gans, "On limits of wireless communications in a fading environment when using multiple antennas," Wireless Pers. Commun., vol. 6, no. 3, pp. 311–335, Mar. 1998.